\documentclass[letterpaper, 10 pt, conference]{ieeeconf}  

\IEEEoverridecommandlockouts                              

\usepackage[utf8]{inputenc}
\usepackage[english]{babel}
\usepackage{graphicx}
\usepackage{amssymb,algorithmic,algorithm}
\usepackage{bbm}
\usepackage{amsmath}
\usepackage[]{cite}

\DeclareMathOperator*{\argmin}{arg\,min}

\newtheorem{assump}{Assumption}

\title{\LARGE \bf
Efficient state and parameter estimation for high-dimensional nonlinear system identification with application to MEG brain network modeling
}

\author{Matthew F. Singh, Chong Wang, Michael W. Cole, and ShiNung Ching
\thanks{\textit{Corresponding Author:} MS is with the Departments of Psychological \& Brain Sciences, and Electrical and Systems Engineering at Washington University in St. Louis, USA and the Center for Molecular and Behavioral Neuroscience at Rutgers University, Newark USA.
{\tt\small f.singh@wustl.edu}}%
\thanks{CW is with the Departments of Psychological \& Brain Sciences, Washington University in St. Louis, USA.
{\tt\small chong.wang@wustl.edu}}
\thanks{MC is with the Center for Molecular and Behavioral Neuroscience at Rutgers University, Newark USA,
{\tt\small michael.cole@rutgers.edu}}
\thanks{SC is with the Department of Electrical and Systems Engineering and Biomedical Engineering, at Washington University in St. Louis, USA,
{\tt\small shinung@wustl.edu}}%
\thanks{MS was funded by NSF-DGE-1143954 from the US National Science Foundation. SC holds a Career Award at the Scientific Interface from the Burroughs-Wellcome Fund. Portions of this work were supported by AFOSR 15RT0189, NSF ECCS 1509342, NSF CMMI 1537015 and 1653589, NSF NCS-FO 1835209 from the US Air Force Office of Scientific Research and the US National Science Foundation, respectively.} 
}

\begin{document}

\setlength{\abovedisplayskip}{2pt}
\setlength{\belowdisplayskip}{2pt}

\maketitle
\thispagestyle{empty}
\pagestyle{empty}

\begin{abstract}
System identification poses a significant bottleneck to characterizing and controlling complex systems. This challenge is greatest when both the system states and parameters are not directly accessible leading to a dual-estimation problem. Current approaches to such problems are limited in their ability to scale with many-parameter systems as often occurs in networks. In the current work, we present a new, computationally efficient approach to treat large dual-estimation problems. Our approach consists of directly integrating pseudo-optimal state estimation (the Extended Kalman Filter) into a dual-optimization objective, leaving a differentiable cost/error function of only in terms of the unknown system parameters which we solve using numerical gradient/Hessian methods. Intuitively, our approach consists of solving for the parameters that generate the most accurate state estimator (Extended Kalman Filter). We demonstrate that our approach is at least as accurate in state and parameter estimation as joint Kalman Filters (Extended/Unscented), despite lower complexity. We demonstrate the utility of our approach by inverting anatomically-detailed individualized brain models from human magnetoencephalography (MEG) data.

\end{abstract}

\section{Introduction}
Control of complex systems benefits greatly from knowledge of the underlying system and the evolution of its states (\cite{Narendra90}), typically in the form of a dynamical systems model. However, in many real-world examples, obtaining such a model is challenging. Even in situations where a general mathematical form of the underlying dynamics is  postulated, a number of unknown parameters typically require specification. The identification of these parameters is often complicated because measurements are opaquely transformed from state variables and/or obfuscated by noise. Substantial progress in system identification research has been made treating these confounds. A wide variety of techniques now enable model-parameterization with well-measured state variables. Conversely, Bayesian methods (e.g., Kalman and Particle filtering  \cite{Kalman,EKF,UKF}) are now well-established for estimating (latent) state variables from measurements with a known system model (state estimation). However, the estimation of both states and parameters (dual-estimation) remains an unmet challenge, especially for large-scale problems relevant to applications in network inference \cite{Sun08}. In particular, dual-estimation problems arise when state-variables are not directly measured or undergo a high degree of mixing en route to eventual measurement. The human brain, for instance, is a high-dimensional nonlinear system in which neither the system parameters nor states can be directly measured \textit{in vivo}.

Current techniques for solving dual-estimation problems often represent system parameters as additional `state' variables with null dynamics. This approach enables the use of conventional state-estimation approaches (i.e., Kalman filters \cite{UKF, jEKF}) to perform dual state-parameter estimation. However, these techniques are computationally cost-prohibitive for systems with a large number of parameters. In systems involving many interacting states (e.g., networks), the total number of parameters typically scales quadratically with the state-variables, resulting in filter complexities of $O(n^{4})$ to $O(n^{6})$ in terms of state variables (depending upon how covariances are represented/stored). As a result, identification of many large-scale systems has proven elusive, leading to subsequent bottlenecks in control design \cite{Control}. 

The above challenges are pervasive in biological systems which often contain many unknown parameters (\cite{Pullan10,SourceEEG}). As a result, model-based control approaches often resort to generic or abstract parameterization, or examination of structural rather than dynamical aspects of control.  In the current work, we propose a simple, but computationally efficient paradigm to estimate the parameters of large nonlinear systems. We verify that our technique performs competitively with current gold-standards (joint Kalman filter, jKF) within the latter's tractable range. We then demonstrate efficacy in solving very high-dimensional problems in simulation and with human magnetoencephalography (MEG) data, with state spaces that would be intractable for jKFs.

\section{Problem Formulation}
\subsection{Dual estimation problem}
We address the problem of estimating parameters for large nonlinear systems in the presence of imperfect state measurement. We consider discrete-time systems of state-variables $x_{t}\in\mathbb{R}^{n}$ evolving according to the nonlinear dynamics $f_{t+1}(x_{t},\theta)$ with process noise $w_{t}\sim\mathcal{N}(0,Q_{t})$. The vector field $f$ is characterized by a set of unknown parameters $\theta$ and $f$ is allowed to vary in time, hence known inputs are absorbed in $f$:
\begin{equation}
x_{t}=f_{t}(x_{t-1},\theta)+w_{t}
\label{eq:BasicX}
\end{equation}
\begin{equation}
y_{t}=H_{t}x_{t}+v_{t}.
\label{eq:BasicY}
\end{equation}
Here, $y_{t}\in\mathbb{R}^{p}$ represents measurements produced by a linear transformation $H_{t}$ of the state-variables $x_{t}$ and measurement noise $v_{t}\sim\mathcal{N}(0,R_{t})$. A key assumption now follows:
\begin{assump}\label{noiseknown}
We assume that the process and measurement noise covariances ($Q_{t}$,$R_{t}$) are available. 
\end{assump}

Our task is to estimate the system parameters $\theta_{t}$ and states $x_{t}$ given knowledge of $H_{t},Q_{t},R_{t}$ and the general functional form $f_{t}$. 


This problem has been conventionally treated by augmenting the state space to include parameters with stationary dynamics (i.e., $\hat{\theta}_{t|t-1}=\hat{\theta}_{t}+\eta_{t}$). This approach reduces the dual state-parameter estimation problem to one of pure state-estimation, but incurs computational-cost in the process which renders these methods infeasible for systems with many parameters. By contrast, we propose to take the alternate path: reducing the problem to parameter-estimation alone. To do so, we define the least-squares state estimates $\hat{x}_{t}$ as a recursive function of the current parameter estimates, $\hat{\theta}$:
\begin{equation}
\hat{x}_{t|t}(\hat{\theta}):=g(\theta,t)=\argmin_{\bar{x}} E\bigg[\|x_{t}-\bar{x}\|_{2}^{2}\bigg| \mathcal{F}_{t}[y]; \hat{\theta} \bigg]
\end{equation}
with $\mathcal{F}_{t}[y]$ denoting the filtration process generated by $y_{t}$ (i.e., the series of measurements up to time $t$).

By leveraging existing approaches (the Kalman filter) for state-estimation, we reduce the state-estimation component to a direct function of the model parameters and the initial guesses for mean/covariance (i.e., $\hat{x}_{t}:=\hat{x}_{t}(\hat{\theta},\hat{x}_{0},\hat{P}_{0}$)). Under this framework, we propose to identify system parameters ($\theta$) by maximizing the parameter likelihood given observations:
\begin{equation}
\theta=\argmin_{\hat{\theta}}\mathcal{L}\big(\hat{\theta}\big|y_{t}\big)
\label{eq:ThetaCost}
\end{equation}
Problems of this sort have been previously treated in low-dimensional settings using alternating algorithms (e.g. Expectation Maxmization) which separately optimize state-estimates and parameter estimates, while holding the other unknown constant. These algorithms have excelled at identifying the parameters of stationary distributions (e.g., for clustering), but have proven less efficient in identifying dynamical systems due to the larger number of statistical dependencies (i.e., observations are not independent). Alternatively, joint optimization approaches (e.g., the joint Kalman Filters) simultaneously evolve estimates of states and parameters. These approaches leverage the temporal dependency of dynamical systems to only consider the joint distribution of parameters with the system state at each moment. This distribution is then evolved in time through recursive filtering. However, while these approaches are more efficient and stable than alternating algorithms, and thus form the current go-to approach for such problems, they are limited in their ability to handle high dimensional systems with many unknown parameters (e.g. networks).

By contrast, we propose to solve Eq. \ref{eq:ThetaCost} by condensing it to a direct function of the unknown parameters $\theta$ and the observed measurements. We perform this reduction by using a fixed likelihood function (corresponding to the ``true" states/parameters) and expressing the state-estimate as a deterministic function of parameters. We then extract the analytic gradients of this function to perform gradient/Hessian-based optimization.  Hence, we solve for the model that produces the best state estimator/predictor. By using gradient-based optimization, our technique can leverage efficient computational methods and inherits certain guarantees regarding convergence of local minima.

\subsection{Methodological approach}
\label{sec:Meth}

First, we leverage knowledge about the noise distributions (i.e., Assumption \ref{noiseknown}) to remove the dependence of the likelihood function $\mathcal{L}$ on the model parameters $\hat{\theta}$. For the (unknown) true values of parameters ($\theta$) and states ($x_{t}$), observations are distributed:
\begin{equation}
y_{t}\sim\mathcal{N}(f_{t}(x_{t-1}),H_{t}Q_{t}H_{t}^{T}+R_{t})
\end{equation}
Leveraging this relationship, we reduce the maximium-likelihood problem (Eq. \ref{eq:ThetaCost}) to minimizing the Mahalanobis distance (affine to log-likelihood) of prediction errors:
\begin{equation}
\theta=\argmin_{\hat{\theta}}\Omega,\;\;\;\;\Omega:=\frac{1}{N}\sum_{k=1}^{N}z_{k}(\hat{\theta})^{T}M_{k}z_{k}(\hat{\theta})
\end{equation}
\begin{equation}
z_{k}(\hat{\theta})=y_{k}-f_{k}(x_{k-1},\hat{\theta})
\end{equation}
\begin{equation}
M_{k}:=(H_{k}Q_{k}H_{k}^{T}+R_{k})^{-1}
\end{equation}
Secondly, we approximate the true state parameter $x_{k-1}$ via an estimate based upon the current parameter estimate $\hat{\theta}$ and the sequence of preceding observations (the natural filtration $\mathcal{F}_{t-1}[y]$). In general, optimal state-estimation is highly nontrivial for large nonlinear systems. Therefore, we use the optimal estimate (in least-squares error sense) under the following premise: 

\begin{assump}\label{Gaussknown}
1) Gaussianity is preserved under $f$, and 2) linearization of $f$ admits computation of second-order moments (covariances). Further, we restrict ourselves to linear estimators.\end{assump}

 Under these premises, the optimal recursive estimate is given by an Extended Kalman Filter (\cite{EKF}) parameterized according to current state estimates:

\begin{equation}
\hat{x}_{t|t-1}=f_{t}(\hat{x}_{t-1},\hat{\theta})
\end{equation}
\begin{equation}
\hat{P}_{t|t-1}=F'(\hat{x}_{t-1},\hat{\theta})\hat{P}_{t-1}F'(\hat{x}_{t-1},\hat{\theta})^{T}+Q_{t}
\end{equation}
\begin{equation}
S_{t}:=H_{t}\hat{P}_{t|t-1}H_{t}^{T}+R_{t}
\end{equation}
\begin{equation}
K_{t}=\hat{P}_{t|t-1}H_{t}^{T}S_{t}^{-1}; G_{t}:=(I-K_{t}H_{t})
\end{equation}
\begin{equation}
\hat{x}_{t}=G_{t}\hat{x}_{t|t-1}+K_{t}y_{t}
\end{equation}
\begin{equation}
\hat{P}_{t}=G_{t}\hat{P}_{t|t-1}.
\end{equation}
The associated errors ($\hat{z}$) and Mahalanobis distance ($J_{t}$) are:
\begin{equation}
\hat{z}_{t}(\hat{\theta})=y_{t}-H_{t}f_{t}(\hat{x}_{t-1},\hat{\theta});\;\;J_{t}(\hat{\theta})=\hat{z}_{t}^{T}M_{t}\hat{z}_{t}
\end{equation}

which, for a fixed initialization $\hat{x}_{0}$, $\hat{P}_{0}$, is a deterministic function of parameter ($\hat{\theta}$). In later comparisons, we used a fixed multiple of identity for $\hat{P}_{0}$ and small, random values for $x_{0}$. 

 In practice, the Gaussianity and second-order linearization assumptions are violated for nonlinear systems, hence the EKF forms `pseudo-optimal' state-estimates. However, these assumptions are weaker than those of some other dual-estimators (e.g. the joint Extended Kalman Filter \cite{jEKF}) which also extends these assumptions to the joint parameter-parameter and parameter-state distributions. We use the EKF to establish prediction errors as a direct function of parameter estimates ($\hat{\theta}$). We then update parameter estimates based upon the gradients of Eq. \ref{eq:ThetaCost} with respect to $\hat{\theta}$. As states are estimated via forward recursion, the gradients are estimated via regression of the total error $\Omega$ (Algorithm \ref{alg:BackProp}) through estimated states and covariances:
\begin{equation}
\frac{\partial \Omega}{\partial x_{t|t-1}}=G_{t}^{T}\frac{\partial \Omega}{\partial x_{t}}-\frac{2}{N}H_{t}^{T}M_{t}(y_{t}-H_{t}\hat{x}_{t|t-1})
\end{equation}
\begin{equation}
U_{t}:=H_{t}^{T}S_{t}^{-1}y_{t}\bigg[\frac{\partial \Omega}{\partial x_{t|t-1}}\bigg]^{T}
\end{equation}
\begin{equation}
Z_{t}:=\bigg(U_{t}+U_{t}^{T}+2G_{t}^{T}\frac{\partial \Omega}{\partial P_{t}}G_{t}\bigg)F_{t}'
\end{equation}
\begin{equation}
\frac{\partial \Omega}{\partial P_{t-1}}=\frac{1}{2}F_{t}'^{T}Z_{t}
\end{equation}
\begin{equation}
\frac{\partial \Omega}{\partial x_{t-1}}=F_{t}'^{T}\frac{\partial J^{t}}{\partial x_{t|t-1}}+\big\langle Z_{t}P_{t-1},F_{t}''\big\rangle
\end{equation}
\begin{multline}
\big\langle Z_{t}P_{t-1},F_{t}''\big\rangle_{i}:=\sum_{j}\sum_{k}\big[Z_{t}P_{t-1}\big]_{j,k}\frac{\partial^{2} f_{j}}{\partial x^{j}\partial x^{i}}\\
=Tr\bigg(\big[Z_{t}P_{t-1}\big]^{T}\frac{\partial F'}{\partial x^{i}}\bigg)=\bigg[vec(Z_{t}P_{t-1})^{T}\frac{\partial vec(F')}{\partial x}\bigg]_{i}
\end{multline}
\begin{equation}
\frac{\partial \Omega}{\partial \hat{\theta}}=\sum_{t=1}^{N}\bigg[\frac{\partial f_{t}}{\partial \hat{\theta}}^{T} \frac{\partial \Omega}{\partial x_{t|t-1}}+\bigg\langle Z_{t}P_{t-1},\frac{\partial F_{t}'}{\partial \hat{\theta}}\bigg\rangle\bigg]
\end{equation}
The parameter estimates ($\hat{\theta}$) can then be updated according to any gradient/Hessian-based optimization algorithm. In later numerical examples we using Nesterov-Accelerated Adaptive Moment estimation (NADAM, \cite{NADAM}), which is a pseudo-Hessian stochastic-gradient algorithm. Our approach is thus summarized by the following steps for each iteration (Algorithm \ref{alg:BackProp}):

\begin{enumerate}
\item Randomly select initial time(s) $t_{0}$
\item Perform fixed-length EKF based upon parameters $\hat{\theta}$ and store errors $\hat{z}_{t}$
\item Backpropagate error gradients through EKF
\item Update parameter estimates $\hat{\theta}$ using error gradients
\end{enumerate}

\begin{algorithm}

\caption{\\Backpropagated Kalman Filter (1 minibatch). This algorithm performs a single minibatch (iteration) of our approach and these steps are repeated until parameter estimates converge (i.e., a local minimum is found).}
\label{alg:BackProp}

\begin{algorithmic}
\STATE Forward Pass: Filtering and calculating error
\STATE Randomly select $t_{0}$
\FOR{$t=1$ \TO $k$}
\STATE $z_{t}=y_{t}-H_{t}f_{t}(\hat{x}_{t-1})$
\STATE $\hat{P}_{t|t-1}=F_{t}'\hat{P}_{t-1}F_{t}'^{T}+Q_{t}$
\STATE $S_{t}=(H_{t}\hat{P}_{t|t-1}H_{t}^{T}+R_{t})$
\STATE $G_{t}=I-\hat{P}_{t|t-1}H^{T}S_{t}^{-1}H_{t}$

\STATE $\hat{x}_{t}=f_{t}(\hat{x}_{t-1})+\hat{P}_{t|t-1}H_{t}^{T}S_{t}^{-1}z_{t}$
\STATE $\hat{P}_{t}=G_{t}\hat{P}_{t-1}$
\ENDFOR
\STATE Backwards Pass: Accumulating error gradients
\STATE $\partial P_{t}=0, \partial x_{t}=0, \partial \theta=0$
\FOR{$t=k$ \TO $1$}
\STATE $\partial \hat{x}_{t|t-1}=G_{t}^{T}\partial\hat{x}_{t}-(2/k)H_{t}^{T}M_{t}z_{t}$
\STATE $U_{t}=H_{t}S_{t}^{-1}y_{t}\partial \hat{x}_{t|t-1}^{T}$
\STATE $Z_{t}=(U_{t}+U_{t}^{T}+2G_{t}^{T}\partial \hat{P}_{t}G_{t})F_{t}'$
\STATE $\partial \hat{P}_{t-1}=(1/2)F_{t}'^{T}Z_{t}$
\STATE $\partial \hat{x}_{t-1}=F_{t}'^{T}\partial \hat{x}_{t|t-1}+\big\langle Z_{t}\hat{P}_{t-1},F_{t}''\big\rangle$
\STATE $\partial \theta=\partial \theta+\frac{\partial f_{t}}{\partial \hat{\theta}}^{T}\partial \hat{x}_{t|t-1}+\big\langle Z_{t}\hat{P}_{t-1},\frac{\partial F_{t}'}{\partial \hat{\theta}}\big\rangle$
\ENDFOR
\RETURN $\partial \hat{\theta}$
\STATE SGD update for $\hat{\theta}$
\end{algorithmic}
\end{algorithm}
 
\subsection{Relationship to Expectation-Maximization algorithms}
At first glance, these steps bear some resemblance to an expectation-maximization (EM \cite{EM}) approach, which alternates between optimizing states and parameters with the other held constant. However, in the proposed framework, rather than alternately updating state/parameter estimates, we collapse the state-estimation component to leave a function solely in terms of parameters. Thus, from a computational standpoint, we have reduced the problem to parameter estimation without latent variables. Our method treats state estimates as intermediary functions of $\hat{\theta}$ that are produced in evaluating the error function $\Omega(\hat{\theta})$. Thus, our method treats the dual estimation problem (estimating states and parameters) without requiring dual-optimization per se (since states are treated as functions rather than unknowns).

\subsection{Application to large networks}
Our approach is most useful for systems in which the number of unknown parameters scales nonlinearly with the number of state variables. This scenario commonly occurs in systems which feature many potential interactions between pairs of state-variables, as occurs in networks/circuits. Network dynamical systems typically evolve according to linear transformations of local, nonlinear functions. One canonical form for such systems ($x,c \in \mathbb{R}^n,\;A,B\in\mathbb{M}_{n\times n}$) is:
\begin{equation}
x_{t+1} = f(x_{t})=Ax_{t}+B\phi(x_{t}+c)
\end{equation}
with $\phi$ a vector of univariate $C^2$ functions ($\phi_{i}$ only depends on $x^{(i)}$). Denoting $\phi$'s Jacobian at $\hat{x}_{t-1}+c$ as $\Phi'_{t|t-1}$, the corresponding gradients are:
\begin{equation} F'_{t}=\hat{A}+\hat{B}\Phi'_{t|t-1}
\end{equation}

\begin{equation}
\frac{\partial \Omega}{\partial x_{t-1}}=\bigg[(\hat{B}\circ(Z_{t}P_{t}))^{T} \textbf{1}\bigg] \circ \frac{\partial^{2} \phi_{t|t-1}}{\partial x^{2}}+F_{t}'^{T}\frac{\partial \Omega}{\partial x_{t|t-1}}
\end{equation}

\begin{equation}
\frac{\partial \Omega}{\partial \hat{A}}=\sum_{t=1}^{N}\bigg[\frac{\partial \Omega}{\partial x_{t|t-1}}(x_{t-1})^{T}+Z_{t}P_{t-1}\bigg]
\end{equation}

\begin{equation}
\frac{\partial \Omega}{\partial \hat{B}}=\sum_{t=1}^{N}\bigg[\frac{\partial \Omega}{\partial x_{t|t-1}}\phi_{t|t-1}^{T}+Z_{t}P_{t-1}\Phi'_{t|t-1}\bigg]
\end{equation}

\begin{equation}
\frac{\partial \Omega}{\partial \hat{c}}=\sum_{t=1}^{N}\bigg[\frac{\partial \Omega}{\partial x_{t-1}}-\hat{A}^{T}\frac{\partial \Omega}{\partial x_{t|t-1}}\bigg]
\end{equation}

with $\textbf{1}$ denoting the one's vector (used to induce addition over each row) and $\circ$ the Hadamard product (elementwise multiplication). All unspecified terms (e.g. $\partial \Omega/\partial P$) are the same as for the general case. We use this specific case to identify a large, nonlinear network in the next section.

\begin{figure*}[thpb]
	\centering
	\framebox{\parbox{5.0in}{
			\includegraphics[width=0.7\textwidth]{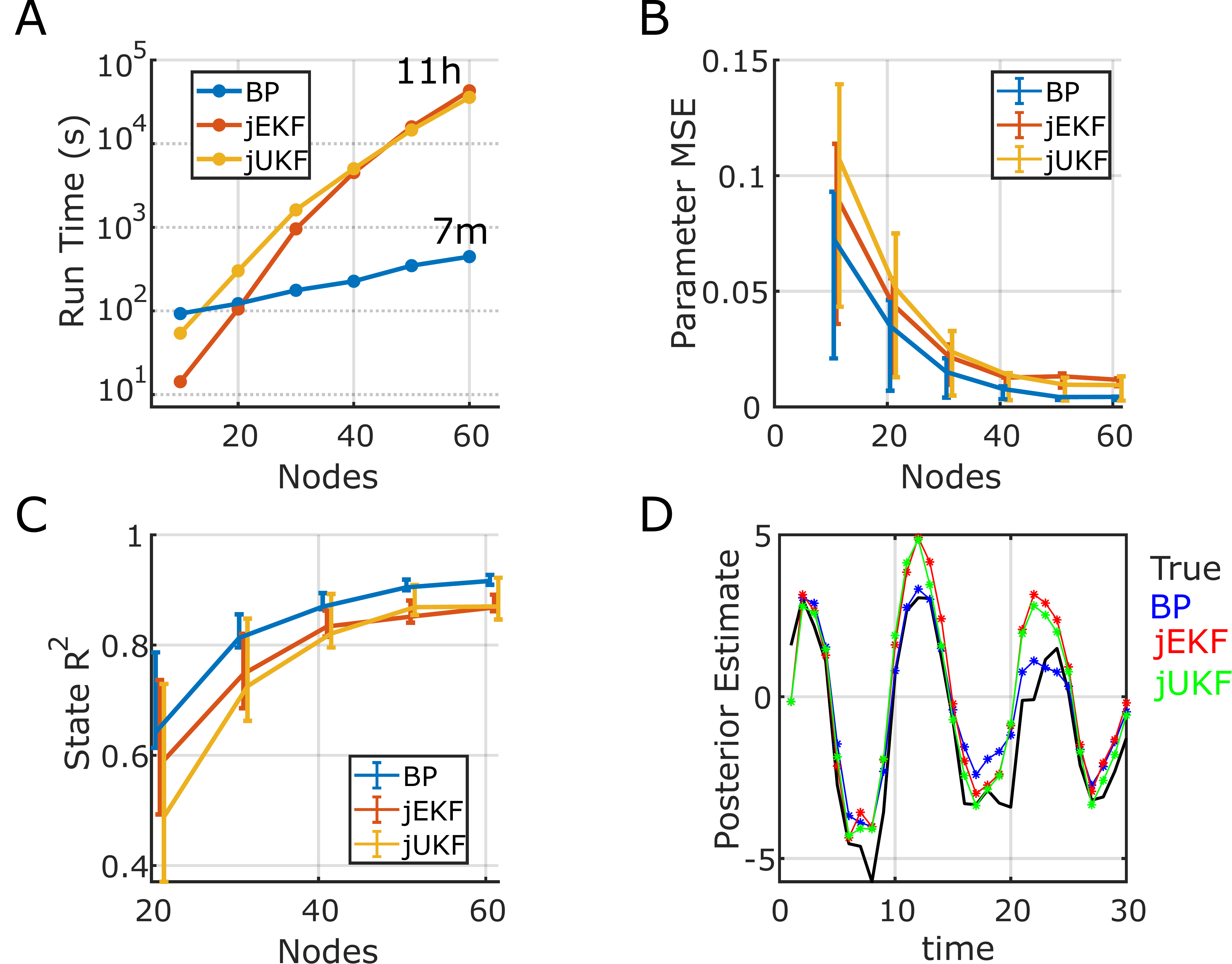}}}
	\caption{The proposed technique is accurate and efficient. A) The technique scales well to large systems and has lower complexity than the joint Kalman Filters. B) Parameter estimates are consistently as-good or better than the joint Kalman Filters across a variety of network sizes. C) State estimation is consistently as-good or better than the joint Kalman Filters. D) Representative filtering time-series (nearest to the mean effect-size) for the first 30 time-points of a 40 node network. Estimated models rapidly converge to accurate state estimates. All state-estimators were initialized identically. We use ``BP" (for backpropagation) to denote the proposed technique. In A-C, lines track mean performance, whereas errorbars indicate the first and third quartiles.}
	\label{fig:Kal}
\end{figure*}

\section{Results}
We view the primary contribution of our method to be enabling dual state-parameter estimation in very large systems. However, we also tested whether the proposed technique is beneficial within the lower-dimensional domains that are applicable to existing methods. We validated our approach in three scenarios: (i) low-dimensional simulations (for benchmarking), (ii) high-dimensional simulations (to validate scalability), and (iii) with high-dimensional data for network discovery and scientific characterization.

\subsection{Identifying complex networks}
For benchmarking, we considered the task of identifying complex networks in the presence of a spatial inverse problem. We generated random recurrent neural networks using the methods in (\cite{Singh20a}) and compared the accuracy and runtime of different approaches to identify the latent states and network connection weights. Recurrent neural networks are a canonical description of nonlinear systems and have been extensively used for benchmarking Kalman-based parameter estimation (\cite{UKF},\cite{Singh20b}). We used networks with the state equation (\cite{Hopfield84}):

\begin{equation}
x_{t+1}=Wtanh(x_{t})+D\circ x_{t}+c+w_{t}
\end{equation}
and rank-reduced measurements $y_{t}=Hx_{t}+v_{t}$.

We also added an additional sparsity constraint to the connectivity matrix: the bottom $40\%$ of connections (entries of $W$, in absolute value) were set equal to zero to ensure that the estimation problem is well-posed. Only non-zero connections were estimated and the local parameters ($D$, $c$) were given. We varied the network size from 10 to 60 nodes. In each case, the measurement space had dimension equal to $40\%$ of the state space. Measurement matrices were randomly generated with singular values distributed $\sim|\mathcal{N}(2,1/4)|$.

We benchmarked performance of our approach against the joint Extended and joint square-root Unscented Kalman Filters (jEKF, jUKF, respectively \cite{jEKF,UKF}). Joint filter parameters were shared between jEKF and jUKF and were manually tuned to maximize performance in parameter estimation using a left-out set of simulations: initial state variance ($P_{0}$) for parameters was $.01$ and parameter process variance was $10^{-5}$ (both iid.). Every 50 timepoints, we symbolically resymmetrized state covariance (ensuring no loss of symmetry over time) and decreased the parameter process covariance by $0.5\%$ to ensure convergence. The total filter length was 30,000 intervals. As per the specification in Section \ref{sec:Meth}, we updated parameters to minimize the Mahalanobis distance: $(y_{t}-Hx_{t})^{T}(HQH^{T}+R)^{-1}(y_{t}-Hx_{t})$ using the NADAM gradient algorithm (\cite{NADAM}) with memory hyperparameters (.98,.95 \cite{Singh20a,Singh20b}) and rate $0.001$. Each of the 125,000 iterations contained 16 filtering/prediction steps. Errors were not calculated during the first 5 steps which was considered a `warm-up' period for state estimation. To promote comparison with joint filters, we only used a single sample per iteration for this analysis, even though our approach enables the use of batch-based updates (i.e. combining across many samples) to parameter estimates. Total examples per network size were: 300 (10 node), 150 (20 node), 75 (30 node), 63 (40 node), 33 (50 node), and 48 (60 node).

Results indicate that the proposed technique is highly scalable (Fig. \ref{fig:Kal}A). In terms of computational complexity, our approach inherits the complexity of the underlying state estimator (EKF) but is not significantly affected by the number of parameters. By contrast, joint-filtering approaches scale nonlinearly with both the number of state variables and the number of unknown parameters (itself a quadratic function of network size). We found that our approach was two orders-of-magnitude faster than jEKF/jUKF in performing dual estimation for 60-node networks. We did find significantly increased run times for the smallest networks considered (10 nodes), thus there may be very low-dimensional scenarios in which the joint-Kalman Filters are preferable. This finding is expected since the proposed technique involves evaluating a large number of Kalman filtrations in terms of state whereas the joint filters require a somewhat smaller number of evaluations for the higher-complexity state+parameter filtration. For large systems, however, results overwhelming favored the proposed technique, which resulted in orders-of-magnitude reduction in run-time.

We found that the proposed technique performed competitively with the joint Kalman-filters in both state and parameter estimation. Estimated parameters were consistently as accurate or better than those obtained using the joint Kalman-filters across model sizes (Fig. \ref{fig:Kal}B). We note that much of the observed variance is attributable to properties of each ground-truth model and the paired-differences between method (i.e. controlling for model) are highly consistent. Accuracy overall improved for all measures as network size increased despite a constant ratio of channels to nodes and constant sparseness of the system matrix. This property was due to the fact that smaller networks were more likely to exhibit poorly-invertible behavior, such as small dynamic ranges due to fixed points in regions which saturate tanh. Larger networks (i.e., $\geq$ 40 nodes) never exhibited this behavior (Fig. \ref{fig:Kal}D). 

We also found that the proposed technique performed at least as well as joint-filters in estimating system states. We used cross-validation for state-estimation in which trained filters were applied to newly generated data from the same model (600 timepoints each). The proposed technique consistently performed at least as well as the joint Kalman filters in performing state estimation (Fig. \ref{fig:Kal}C,D). This benefit is due to more accurate estimates of the underlying system model, since the state estimation component of our approach is identical to EKF. We also tested whether performance depended upon the choice of state-estimation algorithm independent of the algorithm used during the initial dual state/parameter estimation (i.e. using UKF to estimate states based upon the model derived from jEKF). We found that states estimated with UKF were more accurate than those estimated with EKF regardless of which joint Kalman Filter was used during the parameter estimation.

\begin{figure*}[thpb]
	\centering
	\framebox{\parbox{5.0in}{
			\includegraphics[width=0.7\textwidth]{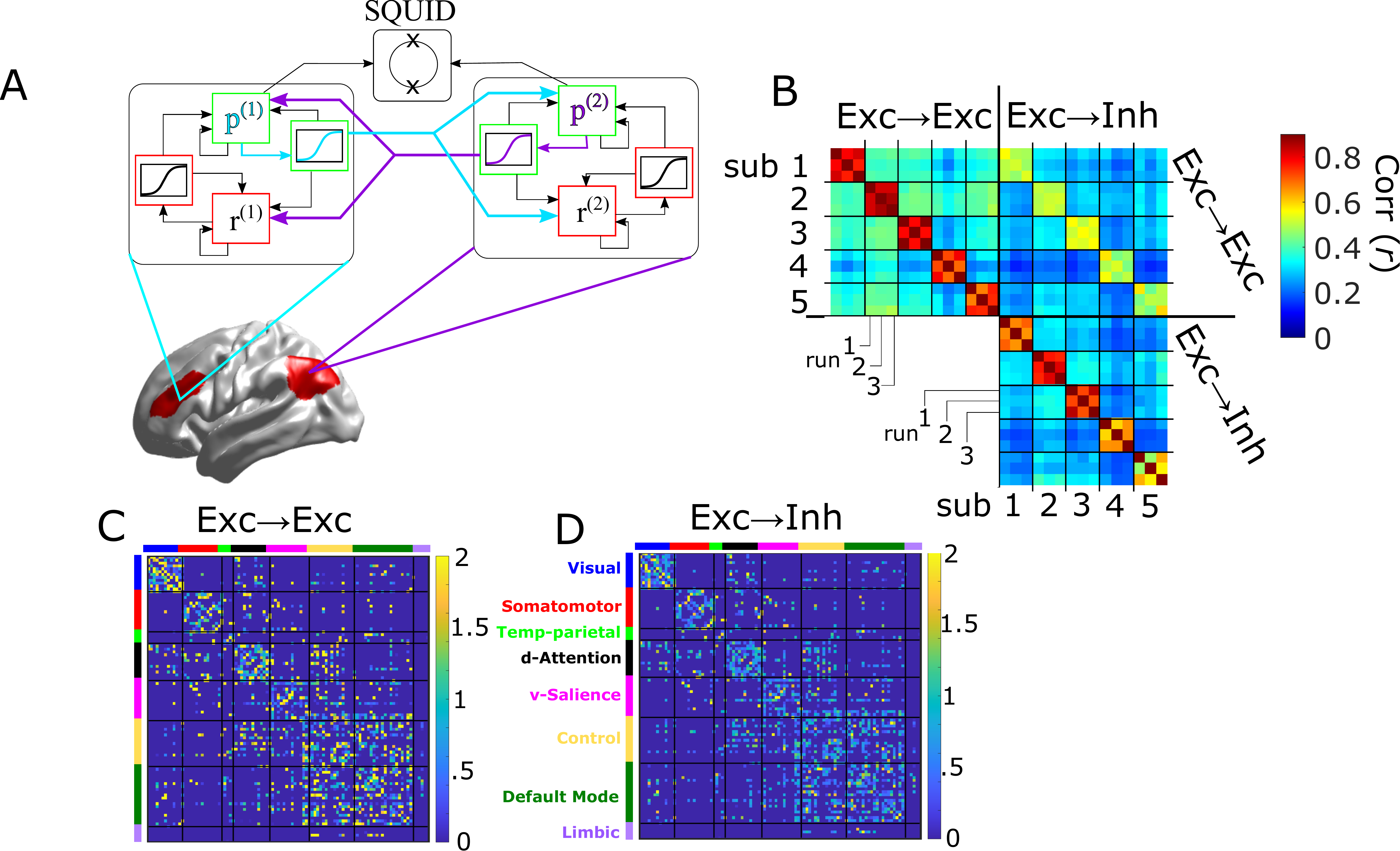}}}
	\caption{A) Block-schematic of the neural system. Each of the 100 brain areas contain interacting excitatory ($p_{i}$) and inhibitory ($r_{i}$) cell populations. Excitatory cells can also connect to other brain areas and generate magnetic fields which are received by a set of magnetometers/gradiometers (SQUID) in the MEG device. B) Individual connectivity estimates from HCP MEG data are highly reliable and subject-specific for both excitatory-to-excitatory and excitatory-to-inhibitory connections. C) Group average E-to-E and E-to-I connectivity estimates.}
	\label{fig:MEG}
\end{figure*}

\subsection{Identification of human brain dynamics}
We demonstrate our method's capability in inferring parameters of a high-dimensional brain network dynamical model from magnetoencelephagraphy (MEG) recordings. We used five (random) subjects' MEG recordings collected during the Human Connectome Project (\cite{HCPorig,HCP}) and which were performed in a shielded room using 248 channel MEG (MAGNES 3600, 4D Neuroimaging, San Diego, CA) sampled at 2KHz. Data was processed according to \cite{HCP} and downsampled to 500 Hz. Each subject contributed three separate ``resting-state" scans lasting 5 minutes each in a single testing session. Due to the spatial separation between the brain and the recording device, the magnetic field experienced by each magnetometer results from many electrical sources distributed across the brain's surface. We calculated the forward model/measurement matrix ($H$) using each subject's structural magnetic resonance image (MRI) and the single-shell boundary element method (\cite{BEM}). We divided the brain into 100 regions (\cite{Schaefer17}) and modeled each region (Fig. \ref{fig:MEG}A) as containing two neuronal populations: excitatory/positive ($p_{t}$) and inhibitory/negative ($r_{t}$):


\begin{equation}
p_{t+1}=\frac{p_{t}}{\tau^{p}}+I_{p}(p_{t},r_{t}) +\omega^{p}_{t}
\end{equation}
\begin{equation}
r_{t+1}=\frac{r_{t}}{\tau^{r}}+ I_{r}(p_{t},r_{t}) +\omega^{r}_{t}
\end{equation}

\begin{equation}
I_{k}^{(i)}=\sum_{j}W^{(k)}_{i,j}\psi_{p}^{(i)}-J^{k}_{(i)}\psi_{r}^{(i)}, \;\;\ k\in\{p,r\}
\end{equation}
Excitatory cells connect to other brain areas (via $W^{p}, W^{r}$), whereas inhibitory cells only connect locally (via $J^{p}, J^{r}$), as compatible with known neuronal micro-circuit physiology (\cite{Micro}). Cells communicate via the sigmoidal function $\psi(x)=tanh(s\circ x+c)$ for $x,s,c\in\mathbb{R}^{n}$. Due to cellular geometry, MEG is only sensitive to excitatory cells which generate systematically oriented current dipoles. Hence, half of $H$ entries were zero. We rank-reduced $H$ to have a maximal matrix condition number of 100 (ratio of largest-smallest singular value) and performed the conjugate dimensionality-reduction on measurements. This resulted in a different dimension of $H$ for each subject/scan (for many scans at least one channel was rejected due to artifact \cite{HCP}), but was typically valued between 70 and 80. Whereas the previous (simulation) analyses only employed a single example per minibatch (to facilitate comparison with jEKF/jUKF), we used 150 examples (starting times) per minibatch for empirical analyses and 50,000 batches. The NADAM rate parameter was lowered to .00025.

Identified brain models were reliable with very similar parameters ($W^{p}$, $W^{r}$) estimated from different scans of the same subject (pairwise-mean $r=.71\pm.06$, $r=.60\pm.11$, respectively). By contrast, parameter estimates had far less similarity between subjects ($r=.35\pm.04$, $r=.24\pm.04$ for $W^{p}$, $W^{r}$, respectively)--indicating the ability to identify individual brain circuitry, as opposed to features which are common among all humans. Moreover, this reliability indicates that our algorithm is converging to a reliable local minimum of the optimization problem for different data drawn from the same non-stationary system (repeating analyses for the same MEG dataset, the algorithm has always converged to the \textit{exact} same minimum).

\section{Conclusion}
We have presented a new technique for efficiently estimating the states and parameters of large nonlinear systems (the ``dual-estimation" problem). This approach consists of substituting a known pseudo-optimal state-estimator (EKF \cite{EKF}) to reduce the dual-estimation problem to parameter estimation. In other words, we solve for the parameters that produce the most accurate EKF predictions. Our results demonstrate that this approach is justified and accurate as well as highly scalable. The primary limitations of our approach are the requirement of prespecified process/measurement noise covariances (Assumption \ref{noiseknown}) and the (soft) requirement that Gaussianity is preserved under the state transition function which can be linearized to compute covariance (Assumption \ref{Gaussknown}). The latter assumption is typically violated, meaning that estimates may be suboptimal and we do not characterize the error relative to the true optimal solution. Future work relaxing these assumptions (e.g., identifying covariances online) will improve generalizability to real-world identification challenges for control.


\addtolength{\textheight}{-12cm}   





\bibliographystyle{unsrt}

\end{document}